\documentclass[dvips]{acta}
\usepackage{supertabular,lscape,epsfig}
\usepackage{amssymb}
\usepackage{amsmath}
\usepackage[T1]{fontenc}

\SetPages{0}{0}

\SetVol{76}{2026}

\usepackage{lmodern}

\newcommand{\BY}{\boldsymbol{y}}
\newcommand{\BZ}{\boldsymbol{z}}
\newcommand{\BW}{\boldsymbol{w}}
\newcommand{\BV}{\boldsymbol{v}}
\newcommand{\BL}{\boldsymbol{\lambda}}
\newcommand{\BA}{\mathbf{A}}
\newcommand{\BB}{\mathbf{B}}
\newcommand{\BC}{\mathbf{C}}
\newcommand{\BD}{\mathbf{D}}
\newcommand{\BM}{\mathbf{M}}
\newcommand{\MA}{\mathcal{A}}
\newcommand{\MD}{\mathcal{D}}
\newcommand{\ME}{\mathcal{E}}

\newcommand{\mFig}[1]{Fig.~\ref{fig:#1}}
\newcommand{\mEq}[1]{Eq.~(\ref{eq:#1})}

\begin{document}

\begin{Titlepage}

\Title{Eigenfrequency sensitivity in nonadiabatic stellar pulsations}

\Author{Zalewski, J.}
{Independent researcher \\
e-mail: jan.zalewski.a2@gmail.com}

\end{Titlepage}

\Abstract{
We derive an adjoint formulation of eigenfrequency sensitivity for linear nonadiabatic stellar pulsations, including the frequency dependence of the boundary conditions. Local sensitivity kernels and their cumulative integrals describe the responses of the excitation rate and oscillation frequency to prescribed scaling of the pulsation matrix and its acoustic, thermal and coupling blocks. Applications to selected radial modes of post-AGB envelopes reveal distinct spatial and subsystem sensitivity profiles. In the cool model, the ordinary mode has weak sensitivity near both boundaries for the slow branch outer boundary conditions, whereas the strange mode exhibits pronounced sensitivity in the outer layers. Selected modes of the hotter model also show appreciable outer-layer sensitivity. The block decomposition reveals differing contributions and cancellations in the excitation-rate and oscillation-frequency responses.

}
{stars: oscillations, stars: AGB and post-AGB, instabilities}

\section{Introduction}

It is long known that in luminous supergiants there appear strange modes (Wood 1976) which can exhibit high excitation rates. There exist also anomalous modes characterized by power integral whose sign is opposite to that of the excitation rate (Zalewski 2026g). These modes differ from ordinary ones in their profiles of eigenfunctions and integral quantities like kinetic energy or power. For such modes, as well as for ordinary ones a question of whether the extent of the assumed pulsation region is adequate appears. It is usually resolved by following the radial profiles of integral quantities (like dissipation integral) to verify that they stabilize near the boundaries of the adopted pulsation region. For high luminosity ($L/M>10^4 L_\odot/M_\odot)$ supergiant stars the radial modes of pulsation enter the asymptotic decaying regime at the bottom of the envelope eliminating the need to model the whole star in order to set up the inner boundary conditions for pulsation which implies placement of the inner boundary at some specified location. At the same time it is well known that outer boundary conditions may affect the pulsations substantially, particularly for strange modes (Zalewski 1992,2026a). For nonradial modes in supergiants, short radial wavelengths and strongly growing or decaying spatial solutions can require the use of WKB approximations (Dziembowski 1977) or numerical methods such as Riccati integration (Glatzel and Gautschy 1992) or Magnus multiple shooting (Townsend and Teitler 2013) opening the questions about the sensitivity of eigenmodes and eigenfrequencies to the assumed treatment of the boundaries and integration methods.

These topics have led to several approaches aimed at examination of the sensitivity of solutions for supergiant stars, particularly with high $L/M$ ratios during their blue-ward or red-ward evolution phases. Shibahashi and Osaki (1981) used a variable parameter to scale the ratio of $\tau_{\rm th}/\tau_{\rm ac}$ in pulsation equations to study how the nonadiabatic terms influence eigenmodes and eigenfrequencies. Zalewski (1992) used homotopy to examine evolution of pulsation spectrum for a post-AGB star with the change of the coupling between thermal and acoustic subsystems. Gautschy and Glatzel (1990) examined the changes in the pulsations of luminous stars by manipulating the thermal time scale. Aikawa (1993) examined changes in linear and nonlinear pulsations by altering the thermo-acoustic coupling as well as the heat capacity scaling. These studies have shown sensitivity of pulsation spectrum of supergiant stars to alterations of the coupling strength between the acoustic and thermal subsystems.

Pulsation modes in supergiants are typically characterized by the eigenmode profiles combined with information resulting from integral quantities following from balance relation (Dziembowski 1977). But as was shown in Zalewski (2026f) the balance relation is computed on the eigenmode and it does not provide direct information about mode eigenfrequency sensitivity. Alternatively and often used are surveys of mode spectra of pulsation for a range of effective temperatures, which show avoided crossings between modes (Saio et al. 1984, Gautschy and Glatzel 1990). 

Reduced models of supergiant envelope pulsation like the toy-model of Zalewski (1992) in which the dynamic subsystem was perturbed by the entropy spike, or models of Buchler et al. (1997) and NAR models of Gautschy and Glatzel (1990) are examples of attempts to decompose the coupled acousto-thermal pulsation problem into smaller units. These studies approach the topic of establishing factors affecting the pulsation mode and frequency using homotopy or toy-models that change the pulsation matrix and thus affect the nature of the pulsation problem for example by transforming it from nonadiabatic to adiabatic or to the NAR limit.

In what follows we propose an approach to the examination of mode sensitivity for nonadiabatic pulsation based on the well established eigenvalue perturbation theory (Greenbaum et al. 2020) and extensively used, in adiabatic formulation, in inverse problems methods (Dziembowski et al. 1990, Christensen-Dalsgaard 2014). Rosenwald and Rabaey (1991) applied an adjoint method to calculate eigenfrequency sensitivities of adiabatic nonradial solar oscillations. Their formulation evaluates derivatives of a characteristic determinant with respect to parameters entering the pulsation equations. We develop our approach for fully nonadiabatic pulsation suitable for high luminosity stars including the effects of boundaries. We then use the sensitivity measure to examine regions of the envelope having most influence on the computed eigenfrequency. In this paper we address the specific question of how the eigenfrequency responds to prescribed changes of the differential equations within the adopted pulsation region.

\section{Adjoint formulation of eigenfrequency sensitivity}
In order to derive a sensitivity measure we start from linear nonadiabatic pulsation equation
\begin{equation}
	\BY'=\BM\BY,
\label{eq:BasicEq}
\end{equation}
where prime denotes derivative with respect to the independent variable $x=\ln(r/R_{\odot})$, $\BY$ is the vector of dependent, linear pulsation variables, and $\BM$ is the pulsation matrix and $\sigma=\gamma+i\nu$ is the nondimensional eigenfrequency. The form of these variables and matrix is the same as in Dziembowski (1977). These equations need to be supplemented by suitable boundary conditions, which in our case  have to be specified at both the inner and outer boundary of the pulsation region. By denoting
\[
\BB_b=\BB(x_b), \quad \BB_s=\BB(x_s),
\]
the boundary conditions may be written as
\[
\BB_b \BY(x_b)=0, \quad \BB_s \BY(x_s)=0,
\]
where the indices denote the inner and outer boundary respectively.

In Zalewski (2026b) the methods used for the solution of the boundary value problem  are discussed. And in Zalewski (2026a) the formulation of boundary conditions is given. 

In what follows we will assume that it is possible to introduce a prescribed perturbation parametrized by its spatial form and of amplitude scaled by $\alpha$. We will apply this perturbation to the matrix $\BM$ and to boundary conditions $\BB$. Thus the boundary value problem in the presence of a perturbation may be stated as
\[
\BY'=\BM(x,\sigma,\alpha)\BY, \quad \BB_b(\sigma,\alpha)\BY(x_b)=0, \quad \BB_s(\sigma,\alpha)\BY(x_s)=0.
\]

The endpoints are kept fixed in space. We assume that the
eigenfrequency $\sigma(\alpha)$ is simple and follows a
differentiable branch. The pulsation matrix $\BM$
and boundary matrices $\BB_e$ are assumed to depend
smoothly on the real parameter $\alpha$ and analytically
on the complex eigenfrequency $\sigma$.

By defining variables 
\[
\begin{aligned}
	\BY_\alpha=\frac{d \BY}{d\alpha}, \quad \sigma_\alpha=\frac{d\sigma}{d\alpha},
\end{aligned}
\]
the derivative of \mEq{BasicEq} may be written as
\begin{equation}
\BY_\alpha'-\BM\BY_\alpha=(\BM_\sigma\sigma_\alpha+\BM_\alpha)\BY,
\label{eq:YAPRIM}
\end{equation}
where $\BM_\alpha=\partial\BM/\partial\alpha$ and $\BM_\sigma=\partial\BM/\partial\sigma$.

The \mEq{YAPRIM} may be simplified by introducing a solution of the adjoint pulsation problem. The adjoint equation may be written as
\begin{equation}
\BZ'=-\BM^\dagger \BZ,
\label{eq:AdjointEq}
\end{equation}
and, as may be shown by direct computation
\[
\frac{d}{dx}\BZ^\dagger \BY=0.
\]
Thus the product of the adjoint and the solution of \mEq{BasicEq} is constant. Our formulation of boundary conditions enforces at the inner boundary $\BZ^\dagger_{x_b} \BY_{x_b}=0$. This relation should also hold then at the outer boundary and this relation may be used as a check of numerical accuracy of integrations - of $\BY$ from the outer boundary inwards to the inner boundary and of $\BZ$ from the inner boundary to the surface. The adjoint solution is obtained using the same integration method as $\BY$.

By combining \mEq{YAPRIM} with \mEq{AdjointEq} it is obtained that
\begin{equation}
	\frac{d}{dx}\left(\BZ^\dagger \BY_\alpha\right)=\BZ^\dagger\left(\BM_\sigma\sigma_\alpha+\BM_\alpha\right)\BY.
\label{eq:SensitivityEq}
\end{equation}

Integrating \mEq{SensitivityEq} over the pulsation region it is obtained that
\begin{equation}
	\left[\BZ^\dagger\BY_\alpha\right]_{x_b}^{x_s}=\sigma_\alpha\int_{x_b}^{x_s}\BZ^\dagger\BM_\sigma\BY dx+\int_{x_b}^{x_s}\BZ^\dagger\BM_\alpha\BY dx.
	\label{eq:SensitivityIntEq}
\end{equation}

To evaluate the left hand side of \mEq{SensitivityIntEq},
we differentiate the boundary conditions with respect to
$\alpha$. At either endpoint $e\in\{b,s\}$ this gives
\begin{equation}
	\BB_e\BY_\alpha(x_e)
	=-\left(\BB_{e,\sigma}\sigma_\alpha
	+\BB_{e,\alpha}\right)\BY(x_e).
	\label{eq:BoundaryDerivative}
\end{equation}
Here $\BB_{e,\alpha}$ denotes the explicit derivative
at fixed $\sigma$, while $\BB_{e,\sigma}$ accounts
for the dependence of the boundary conditions on the
eigenfrequency.

The boundary conditions for the adjoint solution require
$\BZ(x_e)$ to be orthogonal to every vector
admitted by $\BB_e$. Consequently,
$\BZ^\dagger(x_e)$ belongs to the row space of
$\BB_e$ and may be written as
\begin{equation}
	\BZ^\dagger(x_e)
	=\BL_e^\dagger\BB_e,
	\qquad e\in\{b,s\}.
	\label{eq:AdjointBoundary}
\end{equation}
The vectors $\BL_e$ thus express the
endpoint values of the adjoint solution in terms of the
boundary matrices. For boundary matrices of full row rank
they are uniquely determined by $\BZ(x_e)$.

Multiplying \mEq{BoundaryDerivative} by
$\BL_e^\dagger$ and using
\mEq{AdjointBoundary}, we obtain
\begin{equation}
	\BZ^\dagger(x_e)\BY_\alpha(x_e)
	=-\sigma_\alpha
	\BL_e^\dagger
	\BB_{e,\sigma}\BY(x_e)
	-\BL_e^\dagger
	\BB_{e,\alpha}\BY(x_e).
	\label{eq:BoundaryPairing}
\end{equation}
For brevity, denote the two boundary contributions by
\begin{equation}
	Q_e=\BL_e^\dagger
	\BB_{e,\sigma}\BY(x_e),
	\qquad
	P_e=\BL_e^\dagger
	\BB_{e,\alpha}\BY(x_e).
	\label{eq:BoundaryContributions}
\end{equation}
The endpoint term in \mEq{SensitivityIntEq} is therefore
\begin{equation}
	\left[\BZ^\dagger\BY_\alpha
	\right]_{x_b}^{x_s}
	=-\sigma_\alpha(Q_s-Q_b)-(P_s-P_b).
	\label{eq:SensitivityEndpoints}
\end{equation}

Substitution into \mEq{SensitivityIntEq} and collection
of the terms proportional to $\sigma_\alpha$ gives
\begin{equation}
	\frac{d\sigma}{d\alpha}
	=-\frac{
		\displaystyle\int_{x_b}^{x_s}
		\BZ^\dagger\BM_\alpha
		\BY\,dx+P_s-P_b
	}{\MD},
	\label{eq:GeneralSensitivity}
\end{equation}
where
\begin{equation}
	\MD
	=\int_{x_b}^{x_s}
	\BZ^\dagger\BM_\sigma
	\BY\,dx+Q_s-Q_b.
	\label{eq:SensitivityDenominator}
\end{equation}
All quantities are evaluated at the unperturbed solution,
$\alpha=0$ assuming $\MD\ne0$. A related quotient is obtained by Rosenwald and Rabaey (1991, their Eq.~32) by differentiating the pulsation determinant. They evaluate these derivatives using an adjoint of the continuously orthonormalized solution system. Here the sensitivity follows directly from the nonadiabatic pulsation equations and their adjoint, with the derivatives of the boundary conditions appearing explicitly as endpoint contributions.

If the perturbation acts only on the boundary conditions,
then $\BM_\alpha=0$ and
\begin{equation}
	\frac{d\sigma}{d\alpha}
	=-\frac{P_s-P_b}{\MD}.
	\label{eq:BoundaryOnlySensitivity}
\end{equation}
We will not examine this case in the present paper, but it offers the possibility to determine the influence on eigenfrequency coming from perturbations of the boundaries.

Instead, we consider perturbations acting explicitly only
on the pulsation matrix, with
$\BB_{e,\alpha}=0$. Thus the boundary conditions are assumed not to depend explicitly on the perturbation. In this case $P_b=P_s=0$ and
\begin{equation}
	\frac{d\sigma}{d\alpha}
	=-\frac{
		\displaystyle\int_{x_b}^{x_s}
		\BZ^\dagger\BM_\alpha
		\BY\,dx
	}{\MD}.
	\label{eq:MatrixOnlySensitivity}
\end{equation}
The boundary contributions $Q_s-Q_b$ remain in
$\MD$, since the boundary conditions retain
their dependence on $\sigma$. The form of these contributions is derived in the following subsection.

From \mEq{MatrixOnlySensitivity} it follows that it is possible to obtain $d\sigma/d\alpha$ for a suitable perturbation incorporated into the pulsation matrix as $\BM(\sigma,\alpha)$. This enables us to introduce a sensitivity measure, as will be done in the following section.

\subsection{Boundary contributions for conditions based on the local dispersion relation}
It is necessary to obtain explicit forms for the $Q$-terms in \mEq{MatrixOnlySensitivity}. The derived formula for $Q$ is applicable at both boundaries. In what follows we will show that in case of boundary conditions of the form introduced in Zalewski (2026a) an analytic form for $Q$ may be obtained dependent only on eigenvalues and eigenvectors of local dispersion relation as well as on $\BM_\sigma$. To demonstrate this we introduce the following notation
\[
\BM\BV_j=k_j\BV_j, \quad \BW^\dagger_j\BM=k_j\BW^\dagger_j, \quad \BW^\dagger_j\BV_i=\delta_{ji},
\]
where the left eigenvectors are normalized.

In case of the boundary conditions used by us at each boundary there are two admitted vectors and two excluded. We will denote the indices of the admitted vectors by $\MA$ and those of the excluded by $\ME$. These indices are determined by the choice of boundary conditions selectors at each boundary (see Zalewski 2026a). We assume that the selected branches can be followed smoothly under small parameter variations. 

With the adopted description of boundary conditions the boundary matrix $\BB$ consists of the excluded left eigenvectors:
\[
\BB=
\begin{pmatrix}
	\BW_{j_{1}}^\dagger \\
	\BW_{j_{2}}^\dagger
\end{pmatrix}, \quad j_{1}, j_{2} \in \ME.
\]

As a result $\BB\BY=0$ admits the span of the two eigenvectors selected by the boundary selector.

With this notation it is possible to write
\[
\BZ^\dagger=\BL^\dagger \BB = \sum_{j\in \ME}\overline{\lambda}_j\BW_j^\dagger.
\]
Multiplying this equation by $\BV_j$ it is obtained that
\begin{equation}
\overline{\lambda}_j=\BZ^\dagger\BV_j, \quad j \in \ME.
\label{eq:Lambda}
\end{equation}
The $\lambda_j$ are the coefficients of the adjoint vector ($\BZ$) at the boundary in the excluded left-eigenvector basis.

Differentiating the local left-eigenvector equation $\BW_j^\dagger\BM=k_j\BW_j^\dagger$ with respect to $\sigma$ and, subsequently multiplying the resulting equation by the right eigenvector $\BV_i$ it is obtained that:
\[
\left(\partial_\sigma\BW_j^\dagger\right)\BM\BV_i+\BW_j^\dagger\BM_\sigma\BV_i=k_{j,\sigma}\BW_j^\dagger\BV_i+k_j\left(\partial_\sigma\BW_j^\dagger\right)\BV_i.
\]
Since $i\in \MA$ and $j\in \ME$ their biorthogonal product vanishes and therefore
\begin{equation}
\left(\partial_\sigma \BW_j^\dagger\right)\BV_i=\frac{\BW_j^\dagger \BM_\sigma\BV_i}{k_j-k_i}.
\label{eq:Partial}
\end{equation}
We assume that no admitted eigenvalue coincides with an excluded one, $k_i\neq k_j$ for $i \in \MA$ and $j \in \ME$. The solution vector $\BY$ can be written, using left- and right-eigenvectors as
\begin{equation}
\BY=\sum_{i\in \MA}\BV_i\left(\BW_i^\dagger\BY\right).
\label{eq:yvw}
\end{equation}

Using \mEq{Lambda}, \mEq{Partial} and \mEq{yvw} in \mEq{BoundaryContributions} the following expression for $Q$ is obtained
\begin{equation}
 \begin{aligned}
	Q=\BL^\dagger\BB_\sigma\BY&=\sum_{j\in \ME}\overline{\lambda}_j\left(\frac{\partial \BW_j^\dagger}{\partial \sigma}\right)\BY \\
	&=\sum_{j\in\ME}\sum_{i\in\MA}\overline{\lambda}_j\; \left(\frac{\partial \BW_j^\dagger}{\partial \sigma}\right)\BV_i\;(\BW_i^\dagger\BY) \\
	&=\sum_{j\in\ME}\sum_{i\in\MA}(\BZ^\dagger\BV_j)(\BW_i^\dagger\BY)\frac{\BW_j^\dagger\BM_\sigma\BV_i}{k_j-k_i}.
 \end{aligned}	
\label{eq:eqQ}
\end{equation}

Thus $Q$ at each boundary is obtained from left and right-eigenvectors of local dispersion relation and vectors spanning the allowed solution subspace at the respective boundary as well as on $\BM_\sigma$.

Hence \mEq{MatrixOnlySensitivity} and \mEq{SensitivityDenominator} together with values for $Q_s$ and $Q_b$ provided by \mEq{eqQ} enable the computation of the sensitivity, as will be done in the next section.

\section{Sensitivity to pulsation matrix and subsystem scaling}

\subsection{Sensitivity to pulsation matrix scaling}
In what follows we will apply the following perturbation to the pulsation matrix
\begin{equation}
	\BM(x,\sigma,\alpha)=[1+\alpha f(x)] \BM(x,\sigma),
	\label{eq:UniformFx}
\end{equation}
where the symbol $\BM(x,\sigma)$ denotes the unperturbed pulsation matrix given in \mEq{BasicEq}, and $f(x)$ is a real-valued, piecewise continuous function.

With this form of the perturbation the quantity $\BM_\alpha$ is given by
\[
\BM_\alpha=f(x)\BM.
\]

Defining kernel
\[
K_M(x)=-\frac{\BZ^\dagger\BM\BY}{\MD}
\]
it is obtained from \mEq{MatrixOnlySensitivity} that
\begin{equation}
	\frac{d\sigma}{d\alpha}=\int_{x_b}^{x_s}f(x)K_M(x)\,dx.
\label{eq:KernelKM}
\end{equation}

Using this kernel it is possible to define sensitivity probe $S(x)$ as
\begin{equation}
	S(x)=\int_{x_b}^{x}K_M(\xi)\,d\xi.
	\label{eq:Sensitivity}
\end{equation}

We assume the perturbation function $f(x)$ to be of the form
\begin{equation}
f_x(\xi)=
\begin{cases}
	1, & x_b\leq\xi\leq x,\\
	0, & x<\xi\leq x_s.
\end{cases}
\label{eq:fxxi}
\end{equation}
With this form of $f(x)$ the following relation holds
\begin{equation}
	\frac{d\sigma}{d\alpha}=\int_{x_b}^{x_s}f_x(\xi)K_M(\xi)\,d\xi=\int_{x_b}^x K_M(\xi)\,d\xi=S(x).
\end{equation}
Thus $S(x)$ for this form of $f_x(\xi)$ becomes a cumulative probe of the sensitivity of the eigenfrequency of the boundary value problem to uniform scaling of $\BM$ over a range of $[x_b,x]$, while $K_M(x)$ is the local sensitivity density. Both quantities are complex functions. Their real and imaginary parts determine the sensitivity of the excitation rate ($\gamma$) and of pulsation frequency ($\nu$).

In Eq.~(19) of Zalewski (2026f) we have derived the formulae for the normalized contributions to the real and imaginary parts of the balance relation given by $G_\gamma(x)$ and $G_\nu(x)$ and showed that these functions are not independent and in fact $G_\gamma(x)/G_\nu(x)=\gamma/\nu$ provided the denominators are nonzero. Unlike these normalized cumulative balance diagnostics, the real and imaginary parts of \(S(x)\) need not be proportional. They describe the first-order responses of the excitation rate and oscillation frequency to a prescribed perturbation.

Since $S(x)$ is a cumulative quantity, $S(x_2)-S(x_1)$ is a measure of the sensitivity to perturbations of pulsation matrix over the interval $[x_1,x_2]$, while $S(x_s)$ is the global sensitivity of eigenfrequency to the uniform perturbation of $\BM$.

\subsection{Sensitivity to subsystem scaling}
When analyzing factors influencing mode of pulsation the matrix $\BM$ is often decomposed into blocks 
\begin{equation}
	\BM=
	\begin{pmatrix}
		\BA&\BB\\
		\BC&\BD
	\end{pmatrix}.
\end{equation}
 
Here $\BA$ denotes the dynamic subsystem, described in our case by the relative radial displacement and Lagrangian pressure perturbations $(d,p)$, the block $\BD$ is the thermal block composed of entropy and radiation flux perturbations $(s,f)$. The off-diagonal blocks $\BB$ and $\BC$ represent the couplings of thermal subsystem to dynamic one and dynamic to thermal respectively. 

Perturbations of these matrices were studied by Shibahashi and Osaki (1981) and used for radial pulsations by Saio et al. (1984) who were studying effects of scaling, in our notation, of the form $\alpha M_{4,3}$. With this scaling they were examining the effect of altering the thermal time scale on pulsation modes. For radial post-AGB pulsations Zalewski (1992) considered modification of the form $\epsilon \BB$, which for $\epsilon=1$ corresponds to the unperturbed system, and enables to decrease $\epsilon\to 0$ to obtain correspondence to adiabatic eigenfrequency spectrum. Aikawa (1993) studied a more general case of $(\epsilon,\mu)$ scaling also for radial pulsations of post-AGB stars to scale the thermo-acoustic coupling blocks and the thermal equation. His study enabled him to follow the evolution of pulsation spectrum from adiabatic to nonadiabatic and to the NAR limit. These studies considered nonlinear changes of pulsation eigenfrequency with the change of the control parameters.

With the present formalism it is possible to obtain information about the sensitivity of eigenfrequency to an infinitesimal change by introducing perturbations of a form similar to \mEq{UniformFx} but acting separately on the individual blocks. 

Denoting the blocks of $\BM$ by $\BM_J$, where $J\in\{A,B,C,D\}$ the perturbation of a block is given by
\[
\BM(x,\sigma,\alpha)=\BM(x,\sigma)+\alpha f(x)\BM_J(x,\sigma).
\] 
The matrix $\BM_A$ contains the block $\BA$ in its first and second row and column and zeros elsewhere. Likewise for the other matrices $\BM_J$.

Thus we may introduce block sensitivity density $K_{M,J}$ and block sensitivity probe $S_J(x)$, for $J\in \{A,B,C,D\}$ as
\begin{equation}
	\begin{aligned}
      K_{M,J}&=-\frac{\BZ^\dagger\BM_J\BY}{\MD} \\
      S_J(x)&=\int_{x_b}^{x}K_{M,J}(\xi)\,d\xi.
    \end{aligned}
\end{equation}

With these probes the sensitivities of the blocks of $\BM$ can be individually determined, and thus it is possible to compare the effect on the cumulative sensitivity of eigenfrequency from each of the blocks as well as to determine which of the blocks have largest effect on global sensitivity of the eigenfrequency.

It is easy to see that with our adopted form of the perturbation the following relations between the quantities for the whole matrix $\BM$ and for its blocks hold
\[
K_M=\sum_{J\in \{A,B,C,D\}}K_{M,J}, \quad S(x)=\sum_{J\in\{A,B,C,D\}}S_J(x).
\]

For a fixed unperturbed eigenmode, the sensitivity kernel and its cumulative integral depend linearly on the prescribed matrix perturbation. The same construction can therefore be applied to individual matrix elements \(M_{ij}\), or to any prescribed linear combination of such element-wise perturbations.

The present approach is complementary to the one based on homotopy as it provides sensitivity measures of the pulsation problem and in particular it allows probing of the interior of the pulsation region. This will be done in the next section.

\section{Eigenfrequency sensitivity of radial modes}
The pulsation program described in Zalewski (2026b) was supplemented by an upward integration of the adjoint solution of \mEq{AdjointEq} and the computation of the kernels and probes. For the upward integration of $\BZ$ the two vectors spanning the adjoint allowed subspace determined by the inner boundary conditions (Zalewski 2026a) were chosen as the initial values. Their linear combination is determined at the inner boundary from the left null vector of the pulsation matching matrix. The integrations of both $\BY$ and $\BZ$ were carried out using continuous renormalization with tracking transformation (Zalewski 2026b). 

The model with $L=10^4 L_{\odot}$, $M=0.69 M_{\odot}$ and $\log(T_{\rm eff})=3.8$ from the survey of post-AGB pulsations (Zalewski 2026e) was selected. For the comparison of the sensitivity probes we have chosen the low frequency excited mode of the strange mode pair with $\Im(\sigma)\approx 4.5$ and a damped ordinary mode of comparable frequency $\Im(\sigma)\approx 5.07$. The behavior of sensitivity is also compared to that of selected radial modes at higher effective temperatures.

The amplitudes and power balance relation as well as kinetic energy distribution in the envelope for ordinary and strange radial modes were  analyzed in Zalewski (2026c). The radial profiles of sensitivity probes for these modes that are presented in the following sections may be compared to the profiles of kinetic energy or thermodynamic power discussed in Zalewski (2026c,f).

\subsection{Comparison of sensitivity for ordinary and strange mode}

\begin{figure}[htb]
	\includegraphics{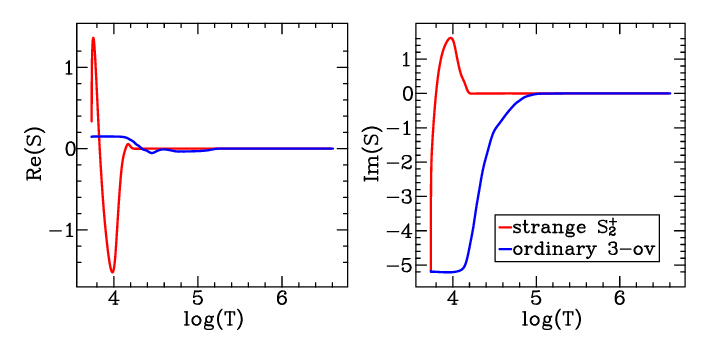}
	\FigCap{Left panel - the $\Re(S)$ is shown for the ordinary 3-ov mode and the strange $S_2^+$ mode. Right panel - the corresponding imaginary parts are shown. It may be seen that the strange mode excitation rate is much more sensitive than for the ordinary mode, while the pulsation frequency $\nu$ shows similar range of sensitivity for both modes.}
	\label{fig:Fig1}
\end{figure}

In \mFig{Fig1} the real and imaginary parts of the probe $S(x)$ are shown for the ordinary, 3-ov mode and for the strange ($S_2^+$) mode.

From \mFig{Fig1} it is seen that for both modes the sensitivity to the changes in $\BM$ deep in the envelope, below the Z-bump zone is small. The regions with $\log(T)> 5$ are practically inert as far as their influence on excitation rate or pulsation frequency is concerned.

Near the surface the sensitivity for the ordinary mode also becomes flat, i.e. the pulsation frequency $\sigma$ in the regions close to surface ($\log(T)<4$) for this mode is not sensitive to the assumed perturbation of $\BM$. However for the strange mode the situation near the surface (regions in and above He~I ionization zone) is different. Both the real and imaginary parts of $S(x)$ for the $S_2^+$ mode change quickly and the change, particularly for the excitation rate is much larger than for the ordinary mode.

This indicates that for the strange mode $S_2^+$ both the excitation rate and mode frequency may be easily affected by changes in $\BM$ in those regions due to the treatment of the outermost layers or outer boundary conditions and their placement. The effects of the adopted form of outer boundary conditions on mode spectrum were discussed in Zalewski (2026a). 

For the ordinary mode the region of largest sensitivity occurs (for this model)  in the He ionization region ($\log(T)\sim 4.4$), with mode frequency showing pronounced sensitivity to changes in $\BM$ as compared to the mode's excitation rate sensitivity. The region of ordinary mode sensitivity is located deeper in the envelope than for the strange mode. This corresponds with the amplitude of the strange modes being much more concentrated towards the surface than for ordinary modes (Saio et al. 1998). In Zalewski (2026f) a comparison of normalized kinetic energy profiles for strange and ordinary mode is shown in Fig.~3.

Wood and Olivier (2014) examined the changes of mode frequency by shifting the outer boundary position and found that the periods of strange modes changed substantially while those of ordinary mode were not so sensitive. This is in agreement with what is seen in the right panel of \mFig{Fig1}. For our models the surface density is $10^{-12} {\rm g/cm^3}$, which, taking into account the optical thickness of the atmosphere above the outer boundary, corresponds to $\tau_{\rm s}\approx 0.05$ for the analyzed model. From \mFig{Fig1} it is seen that at such optical depth the 3-ov ordinary mode pulsation frequency is not affected by conditions near boundary, while for the strange mode it is in this region that the sensitivity is greatest. For the strange mode there is also a large sensitivity of the excitation rate near the surface.

Saio et al. (1998) truncated massive ZAMS models by placing an artificial inner boundary at $\log P=8$. The adiabatic eigenfrequencies of the resulting shallow cavity closely followed the strange-mode frequencies of the complete models, supporting their interpretation of these modes as confined to the outer envelope. For the model in \mFig{Fig1}, the maximum of the approximate critical frequency profile $c_s/(2H_P)$, corresponding to their scaled $\omega_c$, occurs near $\log(T)\approx 5.6$ in the case of our post-AGB models. As may be seen from \mFig{Fig1} this is the region where the eigenfrequency probe $S(x)$ is inert. Therefore, even though Saio et al. (1998) analyzed sensitivity of strange modes to inner boundary placement in massive stars their findings are consistent with the present sensitivity analysis.

\subsection{Ordinary p-modes}
\begin{figure}[htb]
	\includegraphics{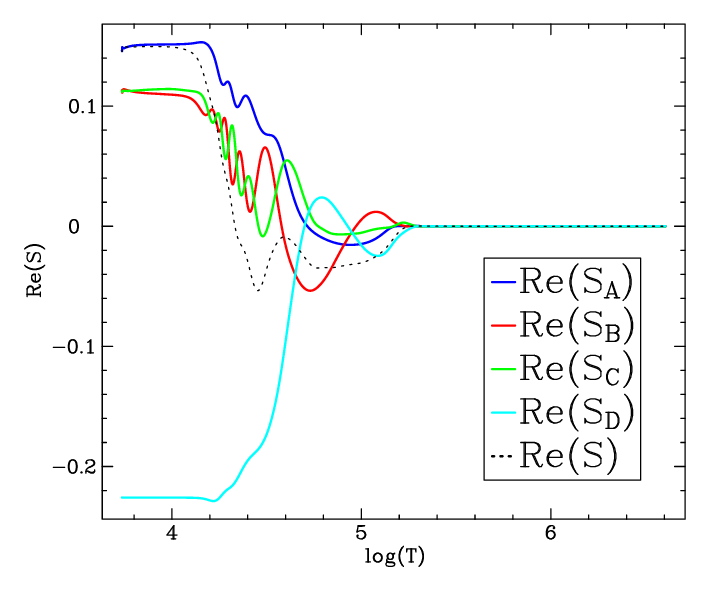}
	\FigCap{The contributions of blocks $\BA$-$\BD$ to the excitation rate ($\gamma$) sensitivity for a 3-ov mode for an envelope model with $\log(T_{\rm eff})=3.8$. As may be seen the real parts of the sensitivity of the blocks as well as of $S$ stabilize for $\log(T)>5.3$ and $\log(T)<4.2$. }
	\label{fig:Fig2}
\end{figure}

As may be seen in \mFig{Fig1} the sensitivity of the excitation rate of the 3-ov mode is much smaller than of the strange mode. The $\Re(S)$ for the overtone mode stabilizes below $\log(T)=4.2$. In \mFig{Fig2} the sensitivity of the excitation rate to scaling of blocks $\BA$-$\BD$ is shown for the 3-ov mode.

It is seen that the sensitivity of the excitation rate to scaling of blocks $\BA$-$\BD$ stabilizes for all four blocks below $\log(T)=4.2$, as does the combined sensitivity $\Re(S)$. For this mode as well as other radial ordinary p-modes for this model the scaling of the $\BD$-block (the main thermal block) decreases the excitation rate while scaling of the $\BA$-block (the main acoustic block) acts in the opposite direction. The proportions of the two blocks vary between ordinary modes. The sensitivity profiles of the coupling blocks $\BB$ (thermo-acoustic) and $\BC$ (acousto-thermal) are similar, but their fluctuations are shifted. The effect of both coupling terms on the excitation rate sensitivity is comparable and occurs in the same direction as for the block $\BA$. Thus blocks $\BA$-$\BC$ counteract block $\BD$ for ordinary modes with the effect that the summed sensitivity $\Re(S(x_s))>0$. The stabilization occurs above the H ionization zone, while in the H and He ionization regions the blocks show variations in sensitivity and stabilize so that the sensitivity of blocks and of $\Re(S)$ becomes near zero in  regions near the Z-bump and deeper in the envelope.

\begin{figure}[htb]
	\includegraphics{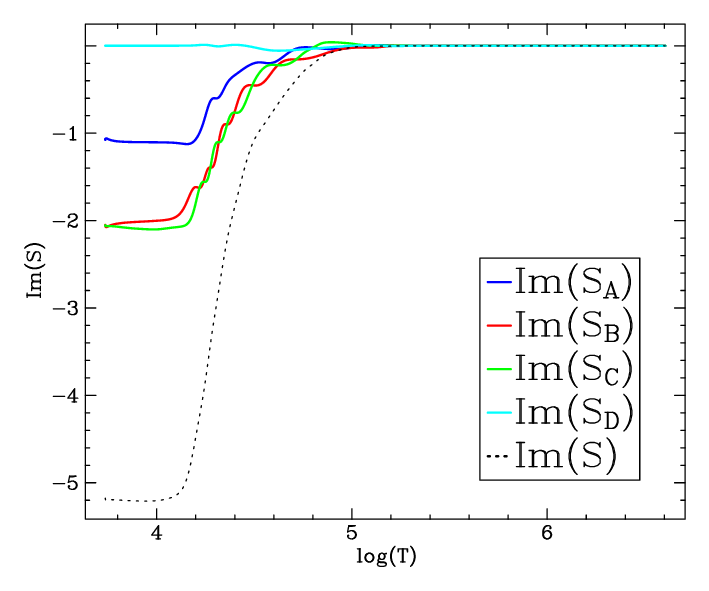}
	\FigCap{The contributions of blocks $\BA$-$\BD$ to the pulsation frequency ($\nu$) sensitivity for a 3-ov mode in a model with $\log(T_{\rm eff})=3.8$. Similarly to the sensitivity of the excitation rate the imaginary parts of the sensitivity of the blocks as well as of $S$ stabilize for $\log(T)>5.3$ and $\log(T)<4.2$. }
	\label{fig:Fig3}
\end{figure}

For the imaginary part of $S$ i.e. for the sensitivity of the pulsation frequency $\nu$ the block sensitivities behave as shown in \mFig{Fig3}.

The sensitivity of $\nu$ due to scaling of the thermal block $\BD$ is very weak $\Im(S_\BD)\approx 0$, while the scaling of block $\BA$ acts to decrease the frequency. The coupling blocks $\BB$ and $\BC$, similarly to the case for the excitation rate sensitivity, exhibit fluctuations shifted along the radial direction. Their effect is negative thus $\Im(S(x_s))<0$, hence scaling of the blocks tends to decrease the frequency of pulsation. Similarly to the case of the excitation rate sensitivity the sensitivity of $\nu$ stabilizes above the H ionization and below the Z-bump. 

The profiles of pulsation frequency block sensitivities for radial ordinary modes are similar. Thus the plots of sensitivity for both $\gamma$ and $\nu$ for these modes show that both deep interior (below Z-bump) and surface regions above H ionization do not have large impact on the sensitivity for the pulsation modes for lower effective temperature models ($\log(T_{\rm eff})<4$). 

\begin{figure}[htb]
	\includegraphics{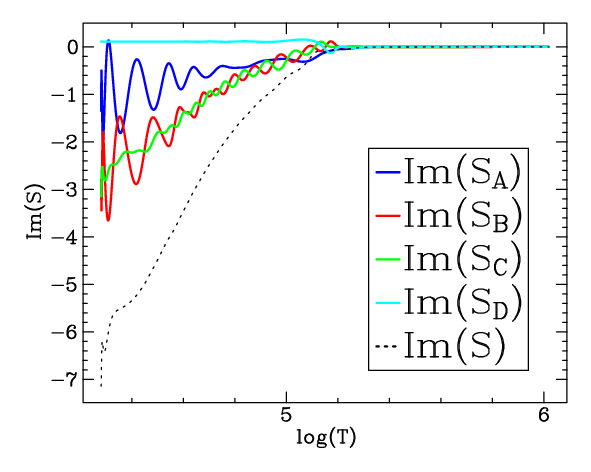}
	\FigCap{The contributions of blocks $\BA$-$\BD$ to the pulsation frequency ($\nu$) sensitivity for a higher overtone radial mode for $\log(T_{\rm eff})=4.35$. While below the Z-bump the sensitivity is close to zero, near the surface the $\Im(S)$ changes substantially, and the block sensitivities $\BA$-$\BC$ exhibit large fluctuations and do not stabilize.}
	\label{fig:Fig4}
\end{figure}

For higher effective temperatures ($\log(T_{\rm eff})>4.3$) the spectra of radial modes become complex, with multiple frequencies and overlapping families of modes (see Fig.~6 in Zalewski 2026f). And while deep in the envelope the radial mode's sensitivity stabilizes, much like for pulsation in colder models, the surface sensitivity does not stabilize. This indicates that the spectra of pulsation in hot models may be sensitive to the treatment of the outer regions of the envelope. This may be seen in \mFig{Fig4} for the sensitivity of $\nu$ for a model with $\log(T_{\rm eff})=4.35$ for a higher order damped mode ($\Im(\sigma)=7.19$). 

For the overtone mode at $\log(T_{\rm eff})=4.35$ the $\Im(S)$ block sensitivities are arranged differently than for the overtone mode at lower effective temperature. In the present case the blocks $\BA$ and $\BB$ show anti-phase correlation along the radius, while block $\BC$ sensitivity follows that of $\BB$ but is much smoother, showing fluctuations only deep in the envelope above the Z-bump. Block $\BD$ sensitivity remains close to zero, so it is similar as for the same block for the colder model. All the block sensitivities combined result in $\Im(S)$ decreasing rapidly in the outer regions of the envelope signifying sensitivity of pulsation frequency to conditions in the regions above He ionization zones. The combined excitation rate sensitivity for this mode likewise does not stabilize near surface.

\begin{figure}[htb]
	\includegraphics{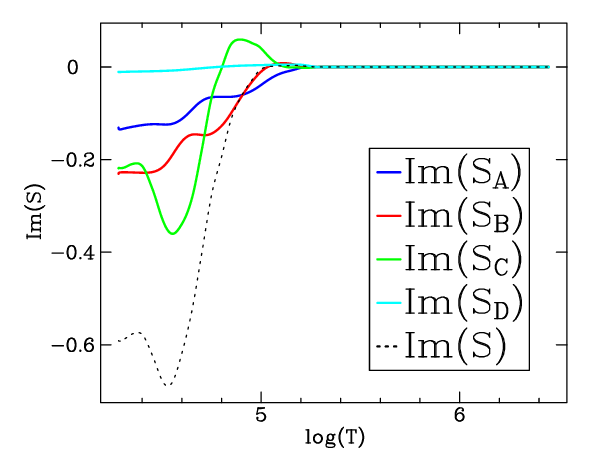}
	\FigCap{The contributions of blocks $\BA$-$\BD$ to the pulsation frequency ($\nu$) sensitivity for a low frequency radial anomalous mode for $\log(T_{\rm eff})=4.35$. The sensitivity of frequency for this mode behaves similarly to the other analysed modes and above the ionization zones its changes are small.}
	\label{fig:Fig5}
\end{figure}

In the case of the anomalous mode ($\Im(\sigma)=0.39$ at $\log(T_{\rm eff})=4.35$) discussed in Zalewski (2026f) and characterized by negative thermodynamic power integral while the mode remains unstable, the sensitivity of the mode near the surface is not so stabilized as for the low temperature mode, but does not exhibit fluctuations which occur for the higher frequency overtone. The plot of $\Im(S)$ for the mode is shown in \mFig{Fig5}.

For the anomalous mode, due in part to its low frequency, the sensitivity tends to stabilize near the surface showing the mode dependence on conditions in the outer envelope but not so pronounced as for the higher overtone shown in \mFig{Fig4}.

\begin{figure}[htb]
	\includegraphics{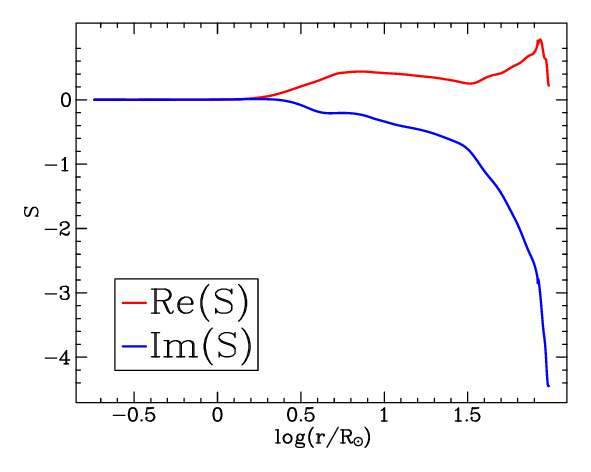}
	\FigCap{The real and imaginary parts of the sensitivity $S$ for an overtone radial mode computed using $(2,4)$ outer boundary conditions. Unlike the case of ordinary mode and $(3,4)$ boundary conditions the sensitivity in case of $(2,4)$ boundary conditions changes substantially in the outer layers. }
	\label{fig:Fig6}
\end{figure} 
 
The sensitivity examples presented so far were for the $(3,4)$ outer boundary conditions selector. This selector leads to high excitation, nearly symmetric (with respect to $\Re(\sigma)$) pairs of strange modes. In Zalewski (2026a) we have shown that the spectrum of modes (ordinary and strange) is sensitive to the form of outer boundary conditions. We have computed the spectrum for the $\log(T_{\rm eff})=3.8$ model using alternative outer boundary conditions $(2,4)$ for which one branch is a fast (large eigenvalue of the local dispersion relation) and one slow. Thus these outer boundary conditions admit the solution subspace at the outer boundary displaying a propagating outward or strongly decaying behavior and one slowly variable in space. The corresponding spectrum may be seen in Zalewski (2026a). In \mFig{Fig6} the real and imaginary parts of the sensitivity $S$ are shown for a mode with $\Im(\sigma)=4.91$, an overtone with a frequency close to the case shown in \mFig{Fig1}. We have plotted $S(x)$ using radial coordinate to show the behavior of sensitivity above the H ionization zones because for these boundary conditions the sensitivity changes substantially near the surface.

\subsection{A strange mode}
\begin{figure}[htb]
	\includegraphics{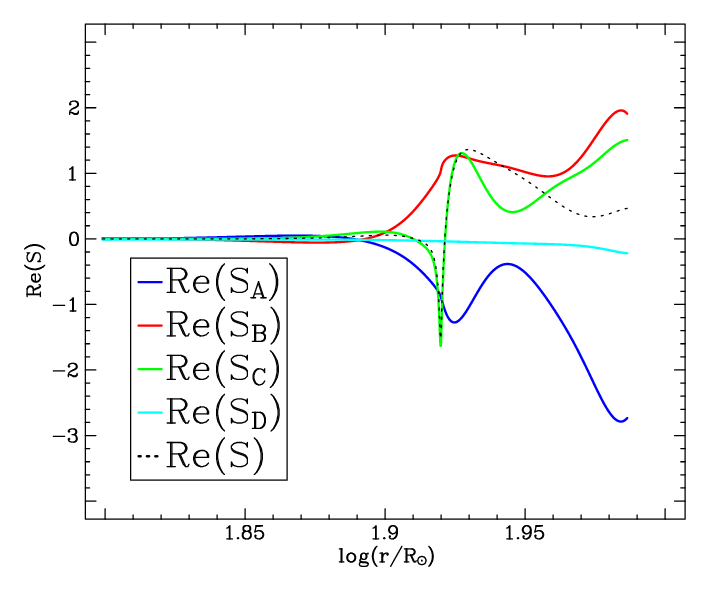}
	\FigCap{The contributions of blocks $\BA$-$\BD$ to the excitation rate ($\gamma$) sensitivity for a $S_2^+$ excited strange mode for an envelope model with $\log(T_{\rm eff})=3.8$. The x-axis is changed to $\log(r/R_{\odot})$ and the displayed range is limited to $\log(r/R_{\odot})>1.8$ to better show the behavior of sensitivity in the outer regions of the envelope. The dip in $\Re(S)$ and $\Re(S_{\BC})$ occurs in the H ionization zone. }
	\label{fig:Fig7}
\end{figure} 
In this section the block sensitivity for a strange mode $S_2^+$ will be analyzed for a model with $\log(T_{\rm eff})=3.8$. 

As may be seen from \mFig{Fig7} the sensitivities for the coupling blocks $\BB$ and $\BC$ for this strange mode behave similarly, however the acousto-thermal coupling block $\BC$ exhibits a sharp fluctuation in the H ionization zone, while the thermo-acoustic $\BB$ stays smooth in this region. The sensitivity of block $\BB$ behaves in anti-phase to the block $\BA$ while the block $\BD$ remains close to zero showing weak sensitivity of excitation rate to the thermal block. The overall excitation rate sensitivity at the surface is smaller than the block sensitivities indicating cancellation.
\begin{figure}[htb]
	\includegraphics{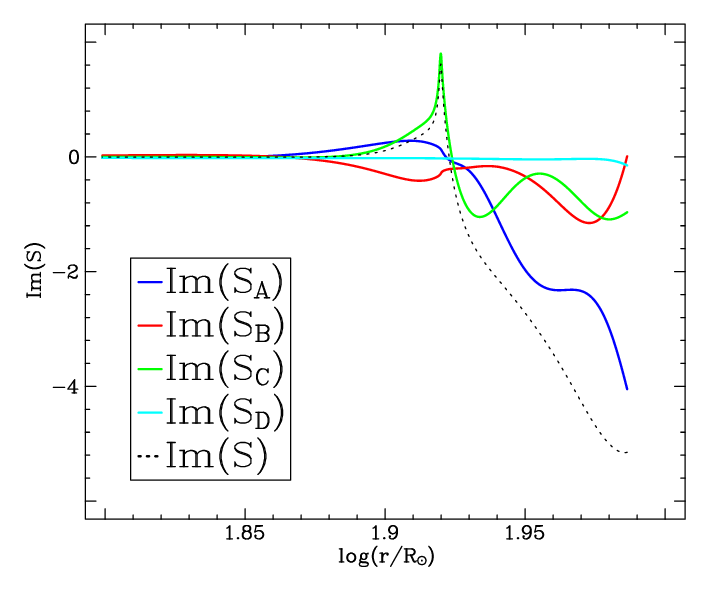}
	\FigCap{The contributions of blocks $\BA$-$\BD$ to the pulsation frequency ($\nu$) sensitivity for a $S_2^+$ excited strange mode for an envelope model with $\log(T_{\rm eff})=3.8$. The frequency is sensitive in the regions above H ionization zone. }
	\label{fig:Fig8}
\end{figure} 

The plot of block sensitivity of the pulsation frequency of the same strange mode is shown in \mFig{Fig8}.

From \mFig{Fig8} it may be seen that the block $\BD$ remains inert while $\Im(S)$ indicates substantial sensitivity in the regions above H ionization zone. The main component of the sensitivity comes from the $\BA$ block, while the coupling blocks exhibit similar and small fluctuations.

Thus for a 3-ov mode the sensitivity of frequency is due to blocks $\BA$-$\BC$ with the coupling blocks playing substantial role, while in case of the strange mode the sensitivity is mainly due to the block $\BA$ with the coupling blocks having smaller influence.

\section{Conclusions}
We have derived eigenfrequency sensitivity measure for nonadiabatic linear pulsations. The formulation is general and enables to obtain information about eigenfrequency sensitivity with respect to any quantity entering the linear nonadiabatic pulsation matrix or the boundary conditions (\mEq{GeneralSensitivity} and \mEq{eqQ}). We then specialize the perturbation to linear scaling of the matrix $\BM$ or its blocks using perturbing functions of the form $1+\alpha f(x)$ (\mEq{UniformFx}) and derive the perturbation kernel ($K_{\rm M}$) and sensitivity probe $S(x)$ (\mEq{Sensitivity}). By assuming a simple form for $f$ given by \mEq{fxxi} it is possible to obtain sensitivity profiles within the pulsation region for a given mode of pulsation. 

The scaling of various blocks of the pulsation matrix using homotopy was analysed in several studies in order to establish changes in pulsation spectrum. Our present approach enables to obtain information about the infinitesimal sensitivity of the eigenfrequency and of the regions of the envelope that contribute most to the sensitivity of the excitation rate or of the pulsation frequency. We have applied this to examine the role of the acoustic, thermal and the coupling acoustic-thermal blocks of the pulsation matrix.

We have analyzed the profiles of the sensitivity for radial ordinary and strange modes in post-AGB envelopes. 

Using our method we have found that for colder envelopes ($\log(T_{\rm eff})<4$) and with the slow-branch outer boundary conditions of $(3,4)$ (see Zalewski 2026a) both the regions below the Z-bump as well as the regions above the H ionization do not influence the eigenfrequency of the ordinary modes substantially. The strange modes exhibit substantial eigenfrequency sensitivity in the outer regions. For hotter envelopes, particularly for $\log(T_{\rm eff})>4.3$ where the spectra of radial pulsations are complex with overlapping families of modes the selected modes exhibit sensitivity in the outer regions, with profiles that differ substantially between modes. 

In our previous study (Zalewski 2026a) we have shown though that the spectra of pulsation for luminous post-AGB envelopes depend on the outer boundary conditions. The insensitivity for ordinary modes near boundaries is obtained here for the $(3,4)$ outer selector. For other selectors, for example $(2,4)$ which admits one fast and one slow branch of local dispersion relation the modes exhibit sensitivity to the outer regions. Thus our present method may be useful for a posteriori examination of the computed eigenmodes.

The block sensitivity analysis shows that ordinary and strange radial modes differ in which blocks contribute most to the eigenfrequency sensitivity. Thus our method provides means of examining the eigenmodes to compare the factors affecting their eigenfrequencies. 

For the outer boundary conditions selector $(3,4)$ the block sensitivity analysis shows that for the cool ordinary modes the excitation rate sensitivities from the acoustic and coupling blocks oppose that of the thermal block. For the pulsation frequency sensitivity for ordinary modes the thermal block has small contribution. For the cool strange mode the acoustic block dominates the frequency sensitivity, while the excitation rate sensitivity exhibits substantial cancellation between block contributions. 

The sensitivity probes identify the envelope regions and pulsation subsystems to which the excitation rate and oscillation frequency respond most strongly. They therefore provide a means of comparing the response of different eigenmodes to prescribed changes of the pulsation equations. This information complements the power-balance diagnostics (Zalewski 2026g) evaluated on the unperturbed eigenmode.


\begin{references}

\refitem{Aikawa, T.}{1993}{MNRAS}{262}{893}

\refitem{Buchler, J.R., Yecko, P.A., Koll\'ath, Z.}{1997}{A\&A}{326}{669}

\refitem{Christensen-Dalsgaard, J.}{2014}{Lecture Notes on Stellar Oscillations, 5th ed, \hfil\break(Aarhus: Aarhus University)}{~}{~}

\refitem{Dziembowski, W.}{1977}{Acta Astron.}{27}{95}  

\refitem{Dziembowski, W., Pamyatnykh, A.A., Sienkiewicz, R.}{1990}{MNRAS}{244}{542}

\refitem{Gautschy, A., and Glatzel, W.}{1990}{MNRAS}{245}{597}

\refitem{Glatzel, W., Gautschy, A.}{1992}{MNRAS}{256}{209}

\refitem{Greenbaum, A., Li, R.-C., Overton, M.L.}{2020}{SIAM Review}{62}{463}

\refitem{Rosenwald, R.D., Rabaey, G.F.}{1991}{ApJS}{77}{97}

\refitem{Saio, H., Baker, N.H., Gautschy, A.}{1998}{MNRAS}{294}{622}

\refitem{Saio, H., Wheeler, J.C., Cox, J.P.}{1984}{ApJ}{281}{318}

\refitem{Shibahashi, H., Osaki, Y.}{1981}{PASJ}{33}{427}

\refitem{Townsend, R.H.D., Teitler, S.A.}{2013}{MNRAS}{435}{3406}

\refitem{Wood, P.R.}{1976}{MNRAS}{174}{531}

\refitem{Wood, P.R., Olivier, E.A.}{2014}{MNRAS}{440}{2576}

\refitem{Zalewski, J.}{1992}{PASJ}{44}{27}

\refitem{Zalewski, J.}{2026a}{submitted to Acta Astron.}{~}{arXiv:2607.19778 [astro-ph.SR]}

\refitem{Zalewski, J.}{2026b}{submitted to Acta Astron.}{~}{arXiv:2607.20979 [astro-ph.SR]} 

\refitem{Zalewski, J.}{2026c}{submitted to Acta Astron.}{~}{arXiv:2608.24515 [astro-ph.SR]}

\refitem{Zalewski, J.}{2026e}{submitted to Acta Astron.}{~}{arXiv:2609.14364 [astro-ph.SR]}

\refitem{Zalewski, J.}{2026f}{submitted to Acta Astron.}{~}{arXiv:2609.34624 [astro-ph.SR]}

\refitem{Zalewski, J.}{2026g}{submitted to Acta Astron.}{~}{arXiv:2610.25166 [astro-ph.SR]}
\end{references}
\end{document}